\documentclass[12pt]{article}
\usepackage{epsfig}
\usepackage{cite}

\begin{document}

\title{CPO prediction: accuracy assessment and impact on UT1 intensive results\footnote{In: D. Behrend, K. D. Baver (Eds.), IVS 2010 General Meeting Proc., NASA/CP-2010-215864, 2010, 261--265.}}
\author{Zinovy Malkin \\ Pulkovo Observatory, St. Petersburg, Russia}
\date{\vspace{-10mm}}
\maketitle

\begin{abstract}
The UT1 Intensives results heavily depend on the celestial pole offset (CPO) model used during data processing.
Since accurate CPO values are available with delay from two to four weeks,
CPO predictions are necessarily applied to the UT1 intensive data analysis, and errors in predictions can influence the operational UT1 accuracy.
In this papers the real accuracy of CPO prediction is assessed using the actual IERS and PUL predictions made in 2007-2009.
Also, results of operational processing was analyzed to investigate the actual impact of EOP prediction errors on the rapid UT1 results.
It was found that the impact of CPO prediction errors is at a level of several microseconds, whereas the impact of
the inaccuracy in the polar motion prediction may be about one order larger for ultra-rapid UT1 results.
The situation can be amended if the IERS Rapid solution will be updated more frequently.
\end{abstract}

\section{Introduction}

Rapid VLBI UT1 observations are vital for accuracy of the rapid IERS EOP solution and its prediction.
To decrease rapid UT1 latency, the special single-base 1-hour sessions are conducted practically every day with delay of processing
from several hours to several days.
As shown in previous studies \cite{Titov2000,Nothnagel2008,Malkin2009} UT1 estimates obtained from the single-base intensive programs
heavily depend on the celestial pole motion model used during analysis.
For the most exacting applications, the celestial pole coordinates are computed as the sum of the theoretical values given by a adopted theory of
precession-nutations, IAU2000A nowadays, and corrections called celestial pole offset (CPO) and obtained from observations, exclusively VLBI nowadays.
The CPO comprises of trends and (quasi)periodic components, Free Core Nutation (FCN) is the first place, caused by the inaccuracy of Earth Rotation theory.

The most accurate CPO can be obtained only from the 24h VLBI sessions and are available, as a rule, with delay from two to four
weeks\footnote{Strictly speaking, CPO results from individual analysis center are available with lower delay, but we consider the IVS combined
CPO series as the most suitable for the EOP service applications}.
Therefore CPO predictions are necessarily applied to the UT1 intensive data analysis, and errors in predictions can influence the rapid UT1 accuracy.
In this papers the real accuracy of CPO prediction is assessed using the actual predictions made by IERS (USNO) and PUL IVS Analysis Center
(Pulkovo observatory).

\begin{figure}[ht!]
\centering
\epsfclipon \epsfxsize=0.75\textwidth \epsffile{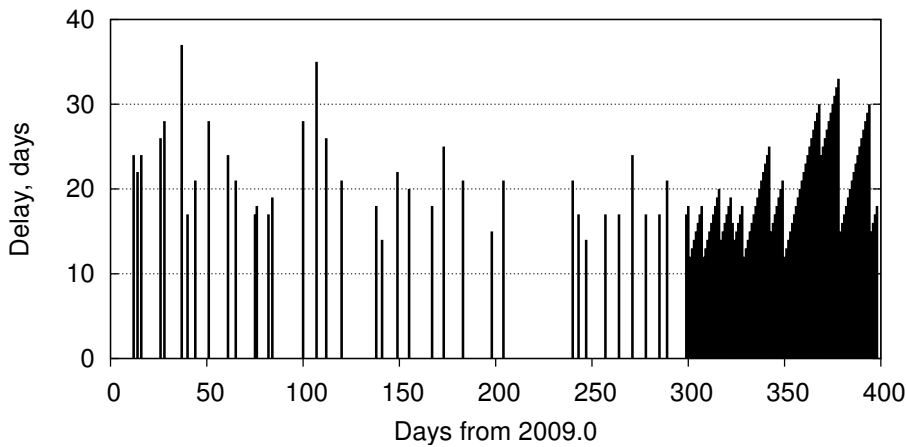}
\caption{Delay of IVS combined EOP.}
\label{fig:ivs_delay}
\end{figure}

The required prediction length can be found from analysis of the IVS combination delay, i.e. the time between the date of publication of
IVS combined solution and the last EOP epoch in this solution (see Fig~\ref{fig:ivs_delay}).
For 2009, the median delay was 18 days, and maximum delay was 37 days.
Of course, IVS series is then updated with new observations processed, but this changes in the IVS data are small enough to significantly
influence rapid UT1 results.
One can see that the required length of CPO forecast is about 40 days.
We extend our analysis to longer length, which may be interesting for other applications.

This paper is aimed at accuracy assessment of the CPO predictions computed with different models.
As usually, the prediction accuracy is derived from comparison of predicted values with the final ones.
For proper interpretation of the results obtained in this study, the following circumstance should be taken into account.
Each CPO model is a result of fitting of observed CPO series.
The models differ not only by method of fitting, but also by CPO data used for analysis, which makes results of accuracy assessment somewhat ambiguous.
One may consider the prediction accuracy with respect to the model itself, which is, in fact, the accuracy of representation of given model.
However, we are interested in the accuracy of representation of the actual celestial pole motion, which is most important for majority of users.
For this reason we use a comparison of CPO predictions with the IVS combined CPO series, which is intended to be an official standard.

\section{Data used}

In this study we present results of processing of VLBI observations made in the framework of the UT1 intensives IVS observing program with different
delay and different CPO models. The following data were used:
\begin{itemize}
\item INT1 sessions, observed on the workdays on the stations KOKEE (Kk) and WETTZELL (Wz); database is normally available in 2--5 days.
\item INT2 sessions, observed at weekends on the stations TSUKUB32 (Ts) and WETTZELL; database is normally available in 1--2 days.
\item INT3 sessions, observed on Monday on the stations NYALES20 (Ny), TSUKUB32, and WETTZELL; database is normally available in the same day.
\end{itemize}

The following actual and publicly available CPO models were tested:
\begin{itemize}
\item IERS final EOP series computed at the Paris Observatory (C04 series) \cite{IERS_AR_2007}.
  It does not contain prediction and thus is equivalent to zero model for rapid data processing.
\item IERS rapid EOP series computed at the USNO (NEOS model) \cite{IERS_AR_2007}.
  It is constructed from analysis of the NEOS combined CPO series and updated daily.
\item Lambert's FCN series computed at the Paris Observatory (SL model) \cite{Lambert2009}.
  As a matter of fact, it can represent only the FCN contribution to CPO.
  However, this model is recommended by the IERS Conventions (2003) as the substitute for CPO.
  It is constructed from analysis of the IERS combined series C04 and updated every several months.
\item Author's ZM2 model computed at the Pulkovo Observatory \cite{Malkin2007}.
  It is constructed from analysis of the IVS combined CPO series and updated daily.
\end{itemize}

\begin{figure}[ht!]
\centering
\epsfclipon \epsfxsize=0.95\textwidth \epsffile{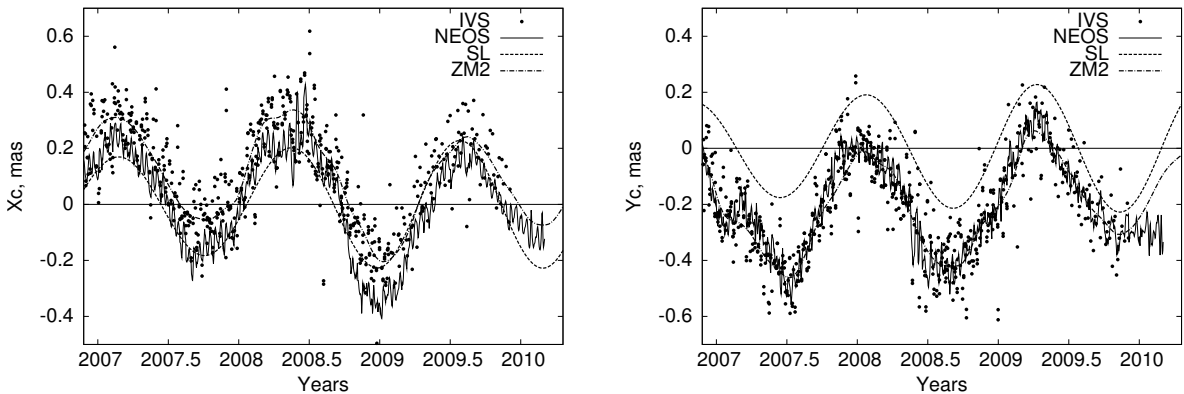}
\caption{CPO models compared with the IVS combined CPO series.}
\label{fig:models}
\end{figure}

Comparison of these models with IVS data is shown in Fig.~\ref{fig:models}.

\section{Accuracy of CPO predictions}

\begin{figure}[ht!]
\parbox{0.48\textwidth}{
\centering
\epsfclipon \epsfxsize=0.48\textwidth \epsffile{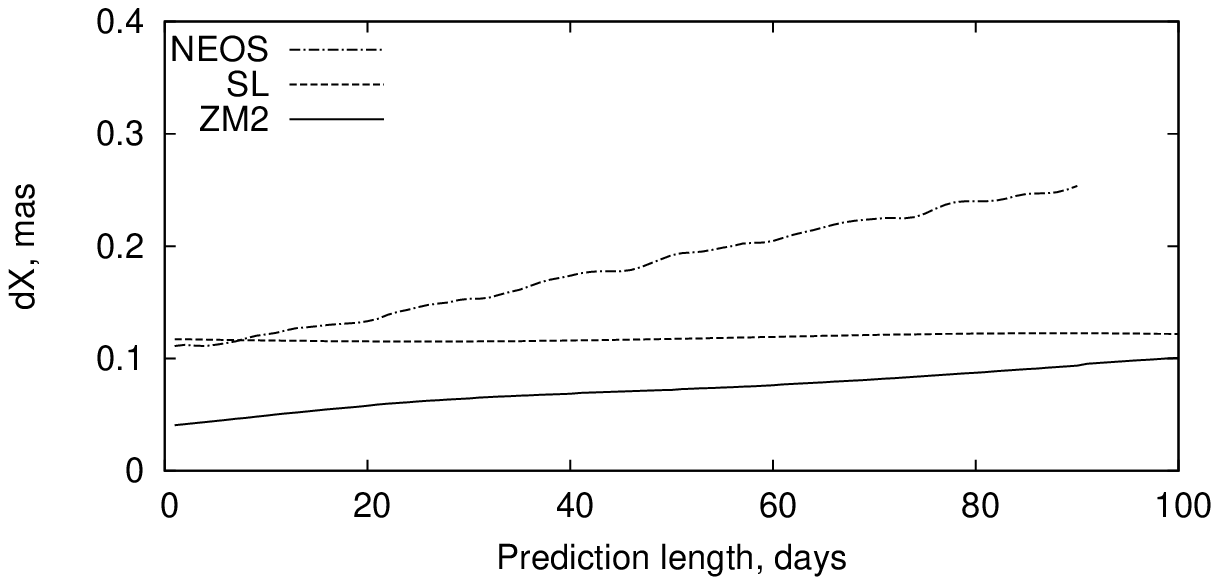} \\
\epsfclipon \epsfxsize=0.48\textwidth \epsffile{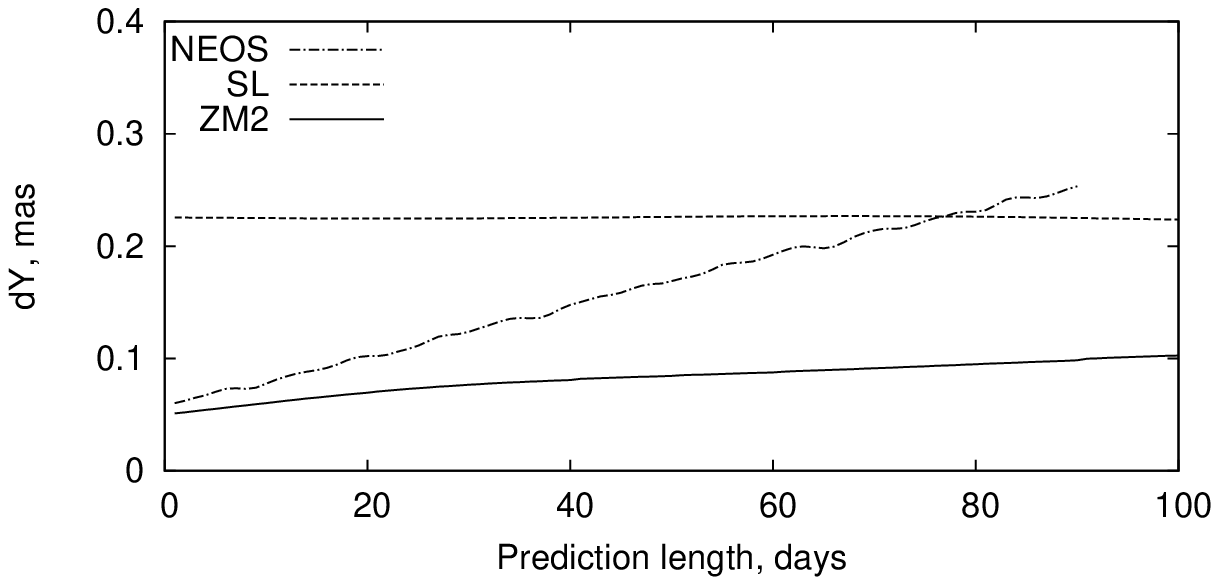}
\caption{The rms error of CPO prediction for different models.}
\label{fig:rms_errors}
}
\hspace{0.03\textwidth}
\parbox{0.48\textwidth}{
\centering
\epsfclipon \epsfxsize=0.48\textwidth \epsffile{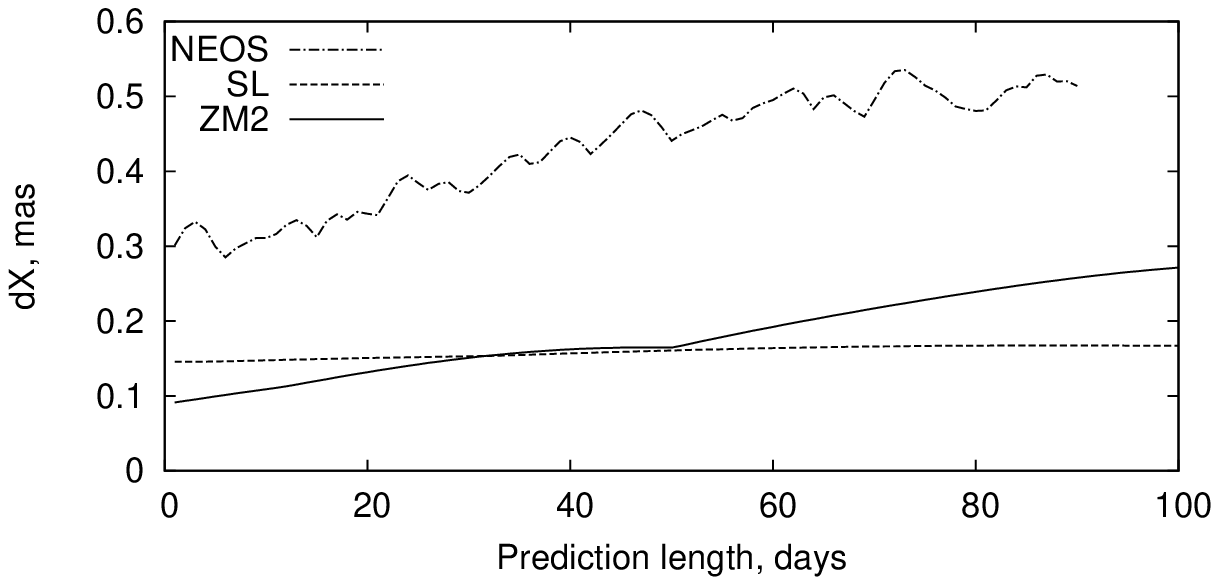} \\
\epsfclipon \epsfxsize=0.48\textwidth \epsffile{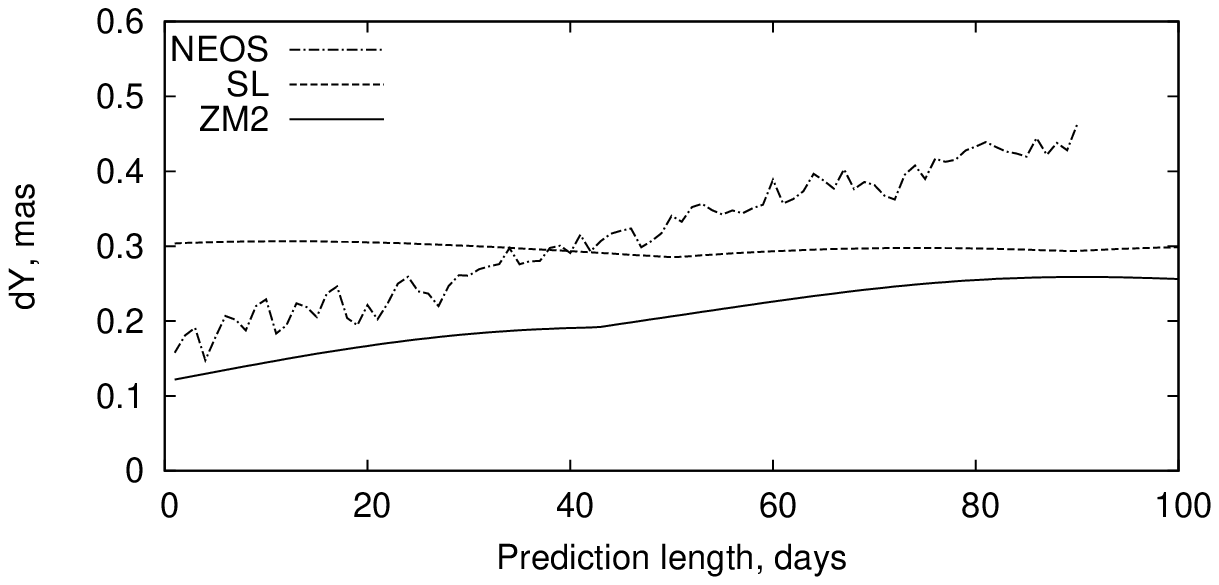}
\caption{The maximum absolute error of CPO prediction for different models.}
\label{fig:max_errors}
}
\end{figure}

As usually, the accuracy of CPO predictions was estimated from comparison of predicted and final values.
Predictions made in the period from December 30, 2006 till December 25, 2009 were used.
Both rms and maximum prediction error was computed; the latter caused the maximum dilution in the UT1 accuracy, and thus is very important.
Results are presented in Figs~\ref{fig:rms_errors} and~\ref{fig:max_errors}.
One can see that ZM2 model provides the best accuracy of CPO prediction.
More details on the accuracy assessment of CPO prediction are given in \cite{Malkin2010}.

\section{Impact of prediction errors on rapid UT1}

To investigate the impact of the CPO prediction error on the rapid UT1 estimates, we started in October 2009 experimental processing
of INT1, INT2, and INT3 sessions with different CPO models.
All the computations presented here were made with the ZM2 model. Using other models gave similar results.
Each session was processed triply:
\begin{enumerate}
\item immediately after the database is available (O); during INT1 and INT2 processing interpolated polar motion (PM) and extrapolated CPO are used,
  during INT3 processing both CPO and PM are extrapolated;
\item 5-7 days after the date of observations (O2); in this case practically final PM is available, but CPO is still extrapolated;
\item at least 10 days after the IVS combined CPO for the date is available (F); during this processing practically final CPO and PM are available.
\end{enumerate}

In the beginning of this work, only operational processing (O) was performed. The first O2 processing was made in December 2009.
The differences between the UT1 estimates obtained with different delays for several sessions are shown in Tables~\ref{tab:INT1} and \ref{tab:INT23}.
These results are quite representative for all the computations.
Comparison of F-O, O2-O, and F-O2 differences shows that errors in extrapolated PM coordinates has more significant impact
on the UT1 estimates then CPO prediction errors.
This can be explained by the fact that while the maximum CPO 30-day prediction error during the last three months was about 0.15~mas (for ZM2 model),
whereas the maximum PM error for the IERS Bulletin A 1-day PM prediction during the same period was about 1.7~mas and 0.94~mas for X and Y pole
coordinates respectively.
Consequently, the impact of the PM prediction errors is about one order larger than the impact of the CPO prediction errors.
The latter is at a level of a few microarcseconds, much less than the uncertainty of UT1 estimates.
This result is in good agreement with Nothnagel's estimate of single-base UT1 bias at the level of 1~microsecond for the 40-microarcsecond bias
in CPO or PM \cite{Nothnagel2008}.
One can see that INT3 results obtained for the whole 3-station network NyTsWz are similar to the single-base solutions TsWz for the same sessions.

\section{Conclusion}

The impact of CPO prediction error on the rapid UT1 results seems to be not very significant, much less than the impact of the PM prediction error.
The most rapid UT1 observations of the INT3 observing program that are correlated in the day of observations,
so that the database is normally available before the IERS Rapid combination used as a priori EOP is updated, show sometimes very large bias
up to several tens microseconds as compared with result of processing made after interpolated IERS PM data is published.
The situation can be amended if the IERS Rapid solution will be computed and published more frequently, say every 6 hours,
after the ultra-rapid IGS combination is ready. Such an approach seems to be much more preferable to a user's home-bred combination
of the IERS and IGS data.

\clearpage

\begin{table}
\centering
\small
\caption{Differences between the UT1 estimates obtained with different delays: a case of the KkWz baseline.
  See explanation in text. Unit: microseconds}
\label{tab:INT1}
\def\arraystretch{0.75}
\begin{tabular}{|l|c|rrr||l|c|rrr|}
\hline
Session & Week- & \multicolumn{3}{|c||}{UT1 differences} & Session & Week- & \multicolumn{3}{|c|}{UT1 differences} \\
code   & day &   F-O &  F-O2 &  O2-O & code   & day &   F-O &  F-O2 &  O2-O \\
\hline
I09357 & Wed &  -2.0 &  -2.0 &   0.0 & I09348 & Mon &  -2.2 &  -2.2 &   0.0 \\
I09362 & Mon &  -1.5 &  -1.6 &   0.1 & I09349 & Tue &  -2.1 &  -1.6 &  -0.5 \\
I09363 & Tue &  -1.7 &  -1.5 &  -0.2 & I09350 & Wed &  -0.2 &  -1.3 &   1.1 \\
I09364 & Wed &  -1.5 &  -1.1 &  -0.4 & I09351 & Thu &  -0.8 &  -0.8 &   0.0 \\
I09365 & Thu &  -1.9 &  -1.6 &  -0.3 & I09352 & Fri &  -0.9 &  -0.8 &  -0.1 \\
I10004 & Mon &  -0.3 &  -1.6 &   1.3 & I09355 & Mon &  -1.3 &  -1.3 &   0.0 \\
I10005 & Tue &  -1.8 &  -1.8 &   0.0 & I09356 & Tue &  -1.6 &  -1.6 &   0.0 \\
\hline
\end{tabular}
\end{table}

\begin{table}
\centering
\small
\caption{Differences between the UT1 estimates obtained with different delays: a case of the TsWz and NyTsWz baselines.
  See explanation in text. Unit: microseconds.}
\label{tab:INT23}
\def\arraystretch{0.75}
\begin{tabular}{|l|c|rrr|rrr|}
\hline
Session & Week- & \multicolumn{6}{|c|}{UT1 differences} \\
code   & day &  \multicolumn{6}{|c|}{~}   \\
\cline{3-8}
       &     & \multicolumn{3}{|c|}{TsWz} & \multicolumn{3}{|c|}{NyTsWz} \\
       &     &   F-O &  F-O2 &  O2-O &   F-O &  F-O2 &  O2-O \\
\hline
K09299 & Mon &  -1.6 &       &       &&& \\
K09304 & Sat &  -3.2 &       &       &&& \\
K09305 & Sun &   9.4 &       &       &&& \\
K09306 & Mon &  47.1 &       &       &&& \\
K09311 & Sat &  -3.1 &       &       &&& \\
K09312 & Sun &   1.6 &       &       &&& \\
K09313 & Mon &   8.4 &       &       &&& \\
K09318 & Sat &   1.5 &       &       &&& \\
K09319 & Sun &  -0.4 &       &       &&& \\
K09320 & Mon &   0.3 &       &       &&& \\
K09325 & Sat &  -1.6 &       &       &&& \\
K09326 & Sun &  -1.0 &       &       &&& \\
K09327 & Mon &  20.8 &       &       &&& \\
K09332 & Sat &   0.6 &       &       &&& \\
K09333 & Sun &   5.7 &       &       &&& \\
K09334 & Mon &  -1.6 &       &       &&& \\
K09339 & Sat &  -1.0 &       &       &&& \\
K09340 & Sun &   0.6 &       &       &&& \\
K09341 & Mon &  -0.9 &       &       &&& \\
K09346 & Sat &  -0.5 &   0.5 &  -1.0 &&& \\
K09347 & Sun &   0.2 &   1.6 &  -1.4 &&& \\
K09348 & Mon &  -9.2 &   0.1 &  -9.3 &&& \\
K09353 & Sat &  -1.3 &  -0.6 &  -0.7 &&& \\
K09354 & Sun &   0.2 &   0.0 &   0.2 &&& \\
K09355 & Mon &   3.5 &  -0.6 &   4.1 &&& \\
K09360 & Sat &  -1.0 &  -1.0 &   0.0 &&& \\
K09361 & Sun &   1.7 &  -1.3 &   3.0 &&& \\
K10002 & Sat &  -2.8 &  -2.7 &  -0.1 &&& \\
K10003 & Sun &  -1.2 &  -2.7 &   1.5 &&& \\
K10004 & Mon &   2.9 &  -1.7 &   4.6 &   0.5 &  -2.1 &   2.6 \\
K10011 & Mon &   5.7 &  -3.0 &   8.7 &   6.4 &  -2.6 &   9.0 \\
K10018 & Mon & -15.3 &  -2.8 & -12.5 & -13.3 &  -2.7 & -10.6 \\
K10025 & Mon &  -4.7 &  -2.5 &  -2.2 &  -4.6 &  -2.5 &  -2.1 \\
K10032 & Mon & -11.9 &  -3.2 &  -8.7 & -10.7 &  -3.1 &  -7.6 \\
\hline
\end{tabular}
\end{table}

\end{document}